\documentclass[aip,jcp,preprint]{revtex4-1}
\pdfoutput=1
\usepackage{graphicx}
\usepackage{siunitx}
\usepackage{booktabs}
\usepackage{float}
\usepackage{amsmath}
\restylefloat{table}

\begin{document}
	
\title{Equation of state and critical point behavior of hard-core double-Yukawa fluids}

\author{J. Montes}
\email[]{jmp@ier.unam.mx}
\author{M. Robles}
\email[]{mrp@ier.unam.mx}
\author{M. L\'opez de Haro}
\email[]{malopez@unam.mx}
\affiliation{Instituto de Energ\'{i}as Renovables,  Universidad Nacional Aut\'oinoma de M\'exico \\
	Privada Xochicalco S/N 62580, Temixco, Mor., M\'exico.}

\date{\today}

\begin{abstract}
	A theoretical study on the equation of state  and the critical point behavior of hard-core double-Yukawa fluids is presented. Thermodynamic perturbation theory, restricted to first order in the inverse temperature and having the hard-sphere fluid as the reference  system, is used to derive a relatively simple analytical equation of state of hard-core multi-Yukawa fluids. Using such an equation of state, the compressibility factor and phase behavior of six representative hard-core double-Yukawa fluids is examined and compared with available simulation results. The effect of varying the parameters of the hard-core double-Yukawa intermolecular potential on the location of the critical point is also analyzed using different perspectives. The relevance of this analysis for fluids whose molecules interact with realistic potentials is also pointed out.
\end{abstract}


\maketitle 

\section{Introduction}
\label{intro}

The statistical mechanics approach to fluid systems aims at being able to describe the thermodynamic and structural properties of the system from the knowledge of the interactions of the particles that constitute it.\cite{McQuarrie0} This is indeed a very hard task and it remains in the general case as an unsolved problem. On the other hand, for practical applications, having an analytical equation of state (EOS) for a given fluid, even if only approximate but otherwise reasonably accurate, is very useful. Such an EOS may be derived either empirically or from statistical mechanics, and once more in the latter case only in very few instances this can be achieved. Nevertheless, insights may be gained from the analysis of (maybe highly simplified) model systems that hopefully capture the main characteristics of the interparticle interactions. Among such versatile models, involving only spherical pair interactions, one finds the hard-core multi-Yukawa (HCMY) fluid whose interparticle potential will be specified below. In particular, the hard-core double-Yukawa (HCDY) model has received for a long time, even rather recently, a lot of attention in the literature (see for instance Refs.\  \cite{JM80,GS83,KJ88,RC89,KC96,Tang:1997,AOAS99,AOS99,BU00,Lin,Lin06,S03,S04,Lin04,R93,R89,G04,PA05,Lin06,PA06,K06,AW07,CBLFB07,APER07, YJ08,AIPR08,LHSRWB10,BBC10,KNMT10,KPH10,KRMT11,Kim11,FH11,OAS13,H15} and references therein). This model has been used in connection with real systems including simple neutral fluids \cite{JM80,GS83,KJ88,KNMT10}, chain-like fluids \cite{Kim11,H15}, liquid metals \cite{R93}, charge-stabilized colloids \cite{Lin06}, dilute solutions of strong electrolytes, molten salts and polymer solutions \cite{R89}, dusty plasmas \cite{R93}, microemulsions \cite{KC96}, globular proteins \cite{CBLFB07}, fullerenes \cite{K06} and hydrogen \cite{OAS13}.

One further asset of (HCMY) fluids is that the statistical mechanics derivation is feasible and different approximate EOS for them are already available. In turn, some aspects of their structural properties and the phase behavior arising from such EOS have been analyzed. Already in the 1980's Konior and Jedzredeck\cite{KJ88} pointed out the likely important role of the HCDY fluid as a reference fluid in the application of the thermodynamic perturbation theory for real fluids. As the former authors also stressed, this could eventually materialize only if its properties become as well known as those of the hard-sphere (HS) fluid. With this in mind, in this paper we would like to go one step further. First we will derive yet another (approximate) analytical EOS for the HCMY fluid by using the first order thermodynamic perturbation theory based on the inverse temperature expansion with the HS fluid as the reference fluid. With this EOS, that has a relatively simple form, we will analyze for the HCDY fluid its performance with respect to simulation results in six representative cases and compute the liquid-vapor coexistence curve (including the critical point) in those cases where it occurs. Finally, we will `sweep' a range of values around the ones of the representative six cases in the parameters space of HCDY fluids in order to characterize the link between the critical behavior and the form of the interparticle potential. Given the fact that different sets of values for the parameters in the potential have been fitted for representing realistic potentials, this characterization may serve as a guide and hopefully give us a clue as to the actual critical behavior of real fluids.

The paper is organized as follows. In Sect.\ \ref{Theor} we provide the theoretical framework leading to the derivation of an approximate analytical EOS for the HCMY fluid. This is followed in Sect.\ \ref{HCDYfluid} by the presentation of the results derived for the compressibility factor, the critical constants and the liquid-vapor coexistence curves (when these last two properties are present in the system) for six representative cases of HCDY fluids. An analysis of the effect of the values of the different parameters of the HCDY potential on the location of the critical point of the system is also carried out here. Finally, the paper is closed in Sect.\ \ref{Conclu} with a summary and some concluding remarks.

\section{Theoretical framework}
\label{Theor}

\subsection{ Thermodynamic perturbation theory for the HCMY fluid}

Consider a system defined by a pair interaction potential $\phi
\left( r\right) $ split into a known (reference) part $\phi _{0}\left(
r\right) $ and a perturbation part $\phi _{1}\left( r\right)$. The usual
perturbative expansion for the Helmoltz free energy $A$ to first order in $\beta
\equiv 1/kT$ (where $k$ is the Boltzmann
constant and $T$ is the absolute temperature) leads to\cite{BH67,Mansoori1,Mansoori2,RS,WCS71}

\begin{equation}
\frac{A}{NkT}=\frac{A_{0}}{NkT}+2\pi \rho \beta \int_{0}^{\infty }\phi
_{1}\left( r\right) g_{0}\left( r\right) r^{2}dr+O\left( \beta ^{2}\right) ,
\label{AO1}
\end{equation}
where $A_{0}$ and $g_{0}\left( r\right) $ are the free energy and the radial
distribution function of the reference system, respectively, $\rho$ is the number density and $N$ is the
number of particles.

For the HCMY fluid the reference system may be forced to be a fluid of hard spheres of diameter $d$, {\it i.e.} one sets
\begin{equation}
\phi _{0}\left( r\right) =\left\{
\begin{array}{cc}
\infty , & r\leq d \\
0, & r>d
\end{array}
\right. .
\end{equation}

In this case the Helmholtz free energy to this order is approximated by
\begin{equation}
\frac{A_{MY}}{NkT}\approx \frac{A_{HS}}{NkT}+2\pi \rho \beta
\int_{d}^{\infty }\phi _{MY}\left( r\right) g_{HS}\left( \frac{r}{d}\right)
r^{2}dr,  \label{AO1b}
\end{equation}
where $A_{HS}$ and $g_{HS}\left( r\right)$ are, respectively, the Helmoltz free energy  and the radial distribution function of a hard-sphere fluid and

\begin{equation}
\phi _{MY}\left( r\right) =-\frac{d}{r}\sum_{i=1}^n\left(\epsilon_i e^{-\lambda_i\left(\frac{r}{d}-1\right)}\right)
\label{DY}
\end{equation}

\noindent is the HCMY potential. The parameters  $\epsilon_i$ and  $\lambda_i>0$ ($i=1\cdots n$), determine the actual shape of the potential and we have identified $d$ with the hard-core diameter. Note that, since $g_{HS}(r)=0$ for $0 \leq r < d$,  Eq.\ (\ref{AO1b}) may be
rewritten by introducing the Laplace transform of $\frac{r}{d}g_{HS}\left(
\frac{r}{d}\right) $  as

\begin{equation}
\frac{A_{MY}}{NkT}\approx \frac{A_{HS}}{NkT}- \frac{12 \eta} {T^*} \sum_{i=1}^n
\left(\kappa_i e^{\lambda_i} G(t=\lambda_i)\right),
\label{ALT}
\end{equation}

\noindent where $\kappa_1=1$, $\kappa_i=\epsilon_i / \epsilon_1$ and we have introduced the packing fraction $\eta=\frac{\pi }{6}\rho d^{3}$ and the reduced temperature $T^*=(\beta \epsilon_1)^{-1}$.

The equation of state of the HCMY fluid to first order of the perturbation expansion readily follows from Eq.\ (\ref{ALT}), namely
\begin{equation}
\begin{split}
Z_{MY}=\frac{P}{\rho k T}=\eta \frac{\partial }{\partial \eta }\left[ \frac{A_{MY}}{NkT}\right]_{T,N}= Z_{HS}- \frac{12 \eta} {T^*} \sum_{i=1}^n \left(\kappa_i e^{\lambda_i} G(t=\lambda_i)\right)\\ 
- \frac{12 \eta^2} {T^*} \sum_{i=1}^n \left( \kappa_i e^{\lambda_i} \left(\frac{ \partial G(t=\lambda_i )}{\partial \eta}\right)_T \right).
\end{split}
\label{ZLJ}
\end{equation}
with $P$  the pressure of the hard-core multi-Yukawa fluid and $Z_{HS}$ the compressibility factor of the hard-sphere fluid. Although, one can perform the partial derivation on the right hand side of Eq. (\ref{ZLJ}) explicitly, the resulting expression is not very illuminating and will therefore be omitted. Nevertheless we should stress that the result implied by Eq. (\ref{ZLJ}) is fully analytical.

Note that, at this level of approximation, in order to close the theoretical framework explicit expressions for $g_{HS}(r)$, $A_{HS}$, $G(t)$ and
$Z_{HS}$ are required. These will be provided in the next subsection.

\subsection{The RFA Method for the HS fluid}
\label{rfaM}
Many years ago Yuste and Santos \cite{Bravo1} developed an
analytical-algebraic method to determine  $g_{HS}\left( r\right) $. Their work uses a rational
function aproximation for the Laplace transform of $rg_{HS}\left( r\right) ,$
and so it is referred to as the Rational Function Aproximation (RFA) method.
This method has proved its usefulness to predict very accurate values of $%
g_{HS}\left( r\right) $ in a wide range of densities, even in the metastable
region past the liquid-solid transition \cite{Bravo2}. Furthermore, the same
kind of approach has been successfully adapted and generalized to other
systems. In what follows, we will provide the main steps of the RFA method. For a detailed presentation see Ref. \cite{HYS08}.
For a hard-sphere (HS) system
in the Percus-Yevick (PY) approximation, the Laplace transform of $\frac{r}{d%
} g_{HS}\left( \frac{r}{d}\right) $ was shown by Wertheim \cite{wertheim} to  have an exact expression of the form:
\begin{equation}
G\left( t\right) ={\cal L}\left[ \frac{r}{d}g_{HS}\left( \frac{r}{d}\right) %
\right] =\frac{t}{12\eta }\frac{1}{1-e^{t}\Phi \left( t\right) },  
\label{1}
\end{equation}
where
$\Phi \left( t\right) $ is a rational function given by
\begin{equation}
\Phi _{PY}\left( t\right) =\frac{
	1+S_{1}^{PY}t+S_{2}^{PY}t^{2}+S_{3}^{PY}t^{3}}{1+L_{1}^{PY}t},  \label{2}
\end{equation}
and the coefficients $S_{1}^{PY},$ $S_{2}^{PY},$ $S_{3}^{PY}$ and $%
L_{1}^{PY}$ are well kown analytical functions of  $\eta
$ the label $PY$ denoting the PY\ results.

This result was generalized in Refs.\ \cite{Bravo1} and \cite{Bravo2} by making the assumption that beyond the PY approximation ${\cal L}\left[ \frac{r}{d}g_{HS}\left( \frac{r}{d}\right)\right]$ may be written as
\begin{equation}
\Phi \left( t\right) =\frac{1+S_{1}t+S_{2}t^{2}+S_{3}t^{3}+S_{4}t^{4}}{%
	1+L_{1}t+L_{2}t^{2}}.  \label{3}
\end{equation}
where the (so far unknown) six coefficients $S_{1},S_{2},S_{3},S_{4},L_{1}$
and $L_{2}$ may be evaluated in an algebraic form by imposing the following
two requirements

\begin{enumerate}
	\item  The first integral moment of the total correlation function $%
	h_{HS}\left( r\right) \equiv g_{HS}\left( r\right) \ -1$, {\it i.e.} $%
	\int_{0}^{\infty }rh_{HS}\left( r\right) dr$, must be well defined and non
	zero.
	
	\item  The compressibility factor $Z_{HS}=P_{HS}/\rho kT\equiv 1+4\eta
	g_{HS}\left( d^{+}\right) $, (where $P_{HS}$ is the pressure of the hard-sphere fluid), must be compatible with the isothermal
	compressibility and the radial distribution function, in the sense that on
	the one hand $\chi _{HS}=\left( d\left( \rho Z_{HS}\right) /d\rho \right)
	^{-1}$ and simultaneously $\chi _{HS}=24\eta \left( 1+\int_{0}^{\infty
	}r^{2}h_{HS}\left( r\right) dr\right) $ .
\end{enumerate}

Under these requirements one finds that

\begin{eqnarray}
\ L_{1} &=&\frac{1}{2}\frac{\eta +12\eta L_{2}+2-24\eta S_{4}}{2\eta +1},
\label{4a} \\
\ S_{1} &=&\frac{3}{2}\eta \frac{-1+4L_{2}-8S_{4}}{2\eta +1},  \label{4b} \\
S_{2} &=&-\frac{1}{2}\frac{-\eta +8\eta L_{2}+1-2L_{2}-24\eta S_{4}}{2\eta +1%
},  \label{4c} \\
S_{3} &=&\frac{1}{12}\frac{2\eta -\eta ^{2}+12\eta ^{2}L_{2}-12\eta
	L_{2}-1-72\eta ^{2}S_{4}}{\left( 2\eta +1\right) \eta },  \label{4d}
\end{eqnarray}
and
\begin{equation}
L_{2}=-3\left( Z_{HS}-1\right) S_{4},  \label{5}
\end{equation}

\begin{equation}
S_{4}=\frac{1-\eta }{36\eta \left( Z_{HS}-1/3\right) }\left[ 1-\left[ 1+
\frac{Z_{HS}-1/3}{Z_{HS}-Z_{PY}}\left( \frac{\chi_{HS} }{\chi _{PY}}-1\right) %
\right] ^{1/2}\right] .  \label{6}
\end{equation}

Here, $Z_{PY}=\frac{1+2\eta +3\eta ^{2}}{\left( 1-\eta \right) ^{2}}$ and $%
\chi _{PY}=\frac{\left( 1-\eta \right) ^{4}}{\left( 1+2\eta \right) ^{2}}$
are the compressibility factor and isothermal susceptiblity arising in the
PY theory.

Note that, for a given $Z_{HS}$, the
radial distribution function is given by
\begin{equation}
g_{HS}\left( \frac{r}{d}\right) =\frac{d}{12\eta r}\sum_{n=1}^{\infty
}\varphi _{n}\left( \frac{r}{d}-n\right) \theta \left( \frac{r}{d}-n\right) ,
\label{7}
\end{equation}
with $\theta \left( \frac{r}{d}-n\right) $ the Heaviside step function and
\begin{equation}
\varphi _{n}\left( \frac{r}{d}\right) ={\cal L}^{-1}\left[ -t\left[ \Phi
\left( t\right) \right] ^{-n}\right] .  \label{8}
\end{equation}
Explicitly, using the residues theorem,
\begin{equation}
\varphi _{n}\left( x\right) =- \sum_{n=1}^{4} e^{t_{i}x} \sum_{m=1}^{n}
\frac{A_{mn} \left(t_{i}\right)}{\left(n - m\right) !} x^{n-m} ,
\label{8bis}
\end{equation}
where
\begin{equation}
A _{mn}\left(t_{i} \right) = \lim_{t \rightarrow t_{i}} \frac{1}{\left( m -
	1\right) !} \left(\frac{d}{dt}\right)^{m-1}\left(t-t_{i}\right) t\left[\Phi
\left( t\right) \right] ^{-n},  \label{8bisa}
\end{equation}
$t_{i}$ being the roots of $1+S_{1}t+S_{2}t^{2}+S_{3}t^{3}+S_{4}t^{4}=0$.

To close the problem we will next specify $Z_{HS}$. A particularly simple and yet accurate equation of state for the
HS fluid is the one due to Carnahan and Starling (CS)\cite{CS} which yields the following expression for $Z_{HS}$
\begin{equation}
Z_{HS}=\frac{1+\eta +\eta ^{2}-\eta ^{3}}{\left( 1-\eta \right) ^{3}}.
\label{CS}
\end{equation}

For this choice of $Z_{HS}$ it follows that
\begin{equation}
\chi_{HS}= \frac{(1-\eta)^4}{1+4\eta+4\eta^2-4\eta^3+\eta^4}
\label{chiCS}
\end{equation}
and
\begin{equation}
\frac{A_{HS}}{N k_B T}=-1 +\ln \left( \frac{6\Lambda^3}{\pi d^3} \right)+\ln \eta +\frac{4 \eta - 3 \eta^2}{(1- \eta)^2}
\label{freeCS}
\end{equation}
where $\Lambda=\frac{h \sqrt{T^*}}{\sqrt{2 \pi m \epsilon_0}}$ is the de Broglie wavelength with $h$ the Planck constant and $m$ the mass of a particle.

The above choice of $Z_{HS}$ and its associated quantities now allow us to compute approximate values of the thermodynamic properties of the HCMY fluid. The results of such calculations will be presented in the next section for the case of the hard-core double-Yukawa fluid.

\section{Thermodynamic properties of the hard-core double-Yukawa fluid}
\label{HCDYfluid}

\subsection{Compressibility factor and liquid-vapor coexistence}

Now we restrict ourselves to the HCDY fluid. To this end, we consider Eq.\ (\ref{1}) (together with Eqs.\ (\ref
{3}) -- (\ref{freeCS})) and use them in Eqs. (\ref{DY}) -- (\ref{ZLJ}) with $n=2$ and $\kappa_2=\kappa$. In order to test the usefulness of the resulting equation of state for this system, we will compare with the Montecarlo simulation results of Lin {\it et al.} \cite{Lin} who considered the HCDY fluid with both attractive and repulsive interactions outside the core. Specifically, they
examined the following six cases:\\

\noindent case 1: $\lambda_1=1.8$, $\lambda_2=4.0$, $\kappa=1.0$ \\
case 2: $\lambda_1=1.8$, $\lambda_2=2.0$, $\kappa=-1.0$ \\
case 3: $\lambda_1=1.8$, $\lambda_2=4.0$, $\kappa=-3.0$ \\
case 4: $\lambda_1=1.8$, $\lambda_2=8.0$, $\kappa=-6.0$ \\
case 5: $\lambda_1=1.8$, $\lambda_2=8.0$, $\kappa=-12.0$ \\
case 6: $\lambda_1=2.8647$, $\lambda_2=13.5485$, $\kappa=-1.4466$ \\

As the above authors pointed out, the choice $\lambda_1=1.8$ corresponds to the one usually employed in the dispersion interaction between colloidal particles. In case 1, there is no repulsive interaction outside the core, while in case 2 there are both attractive and repulsive interactions, and the total potential is negative for all distances beyond the hard core. Cases 3 -- 6 have repulsive total interaction near the hard core, but the interaction becomes attractive at enough long distances. In cases 4 and 5 the parameters correspond to those that have been used in connection with charged protein molecules, while those of case 6 were proposed by Tang {\it et al.} \cite{Tang:1997} to model a Lennard-Jones fluid.

\begin{figure}
	\includegraphics[height=5cm]{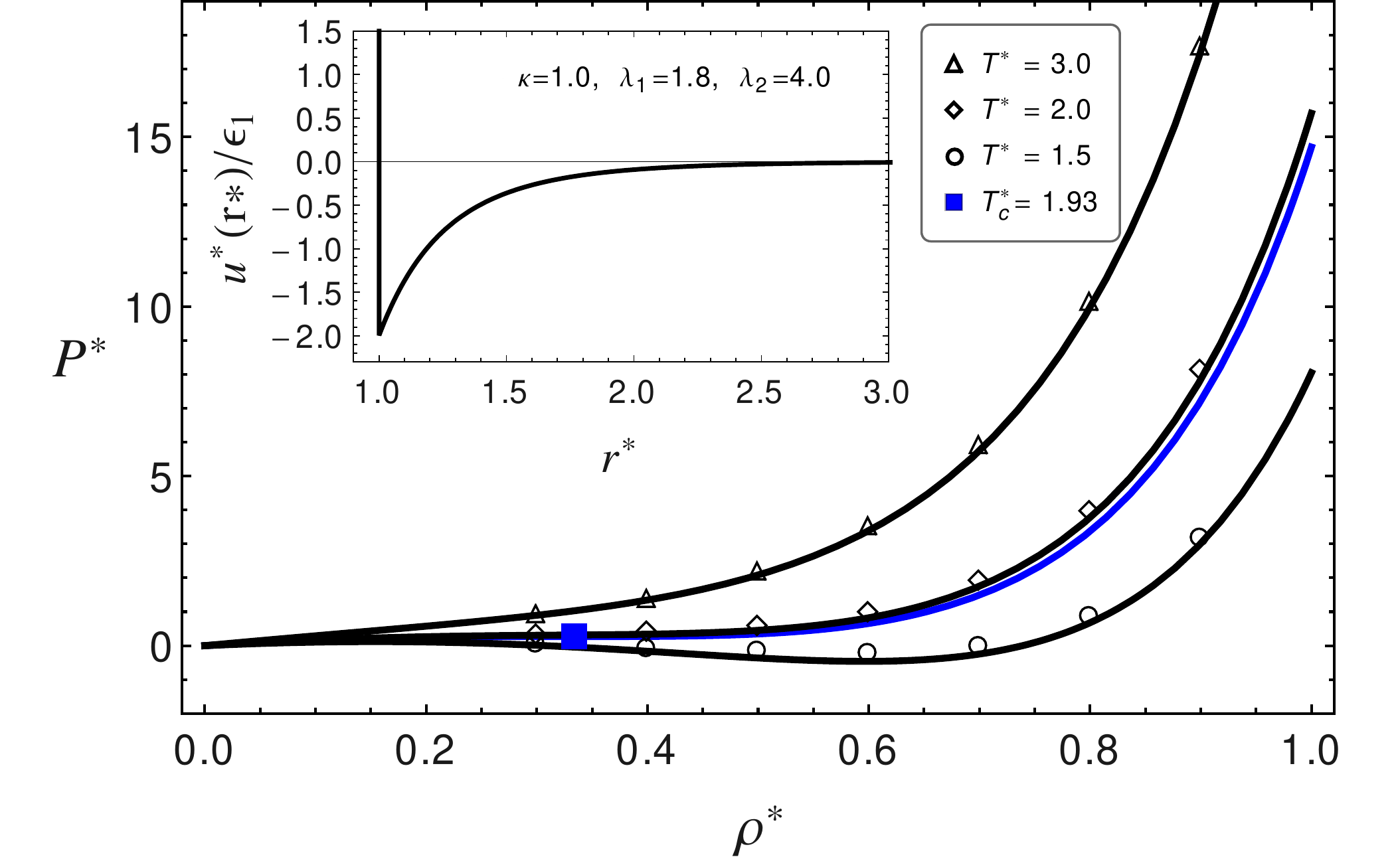}
	\includegraphics[height=5cm]{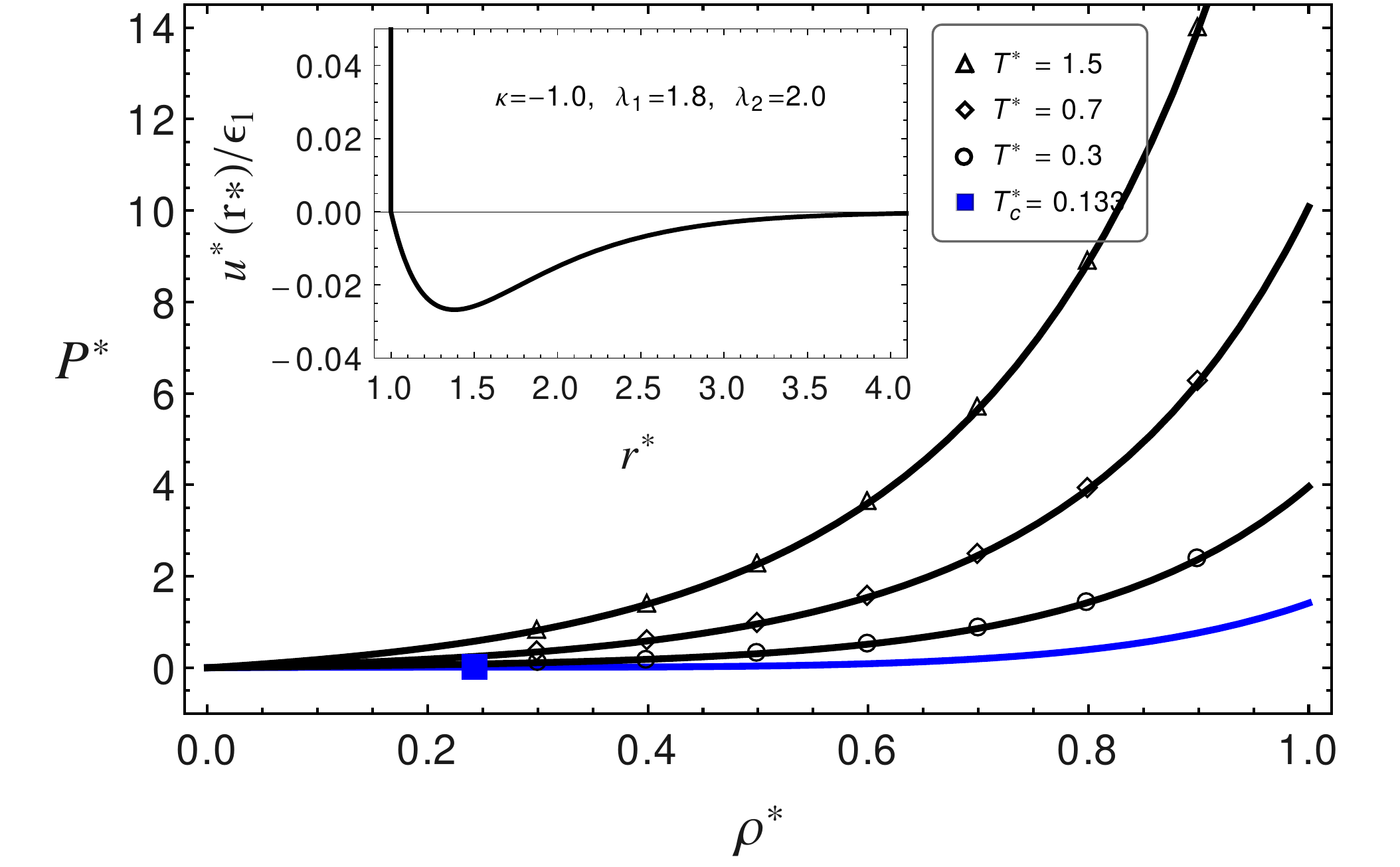}\\
	\includegraphics[height=5cm]{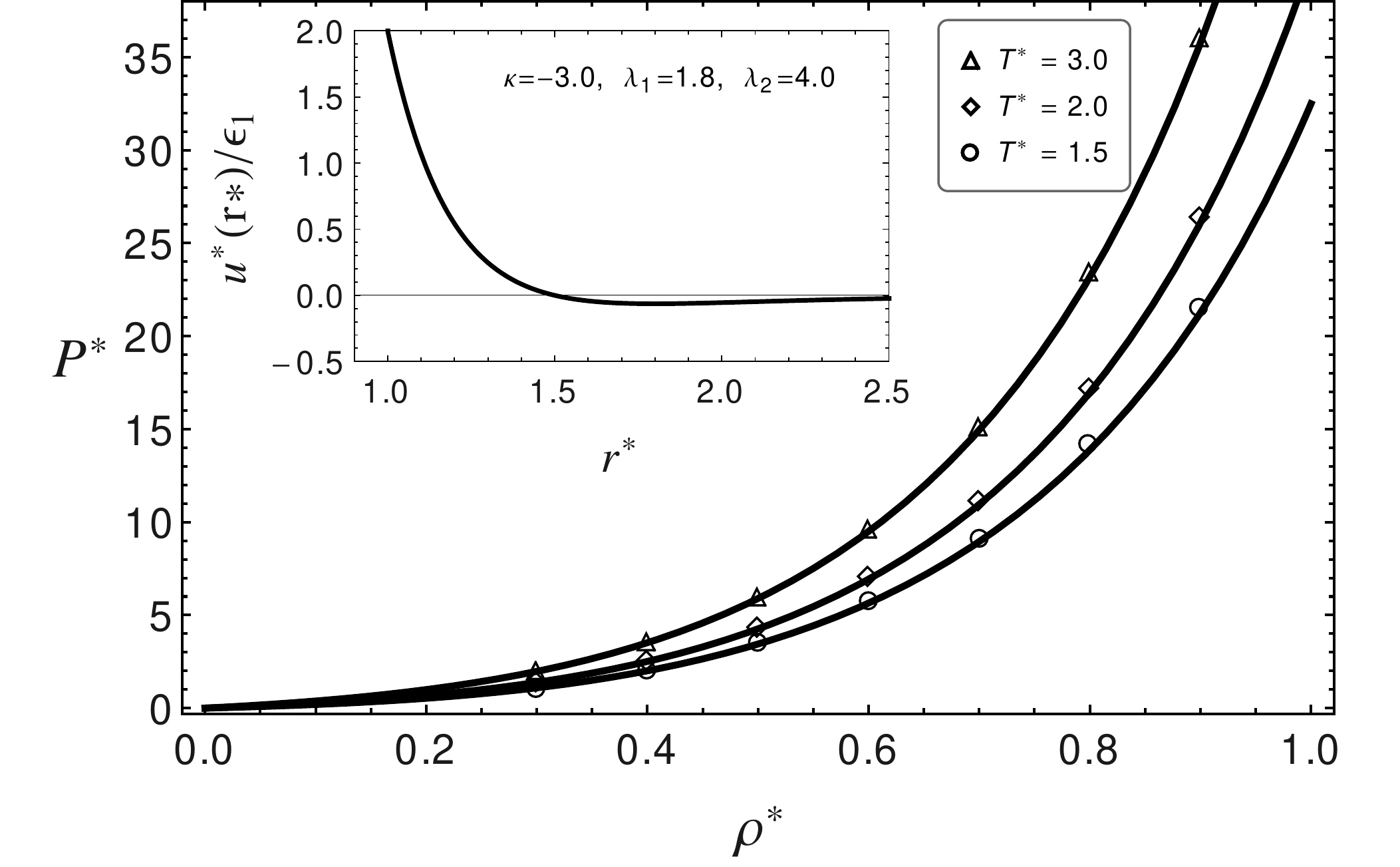}
	\includegraphics[height=5cm]{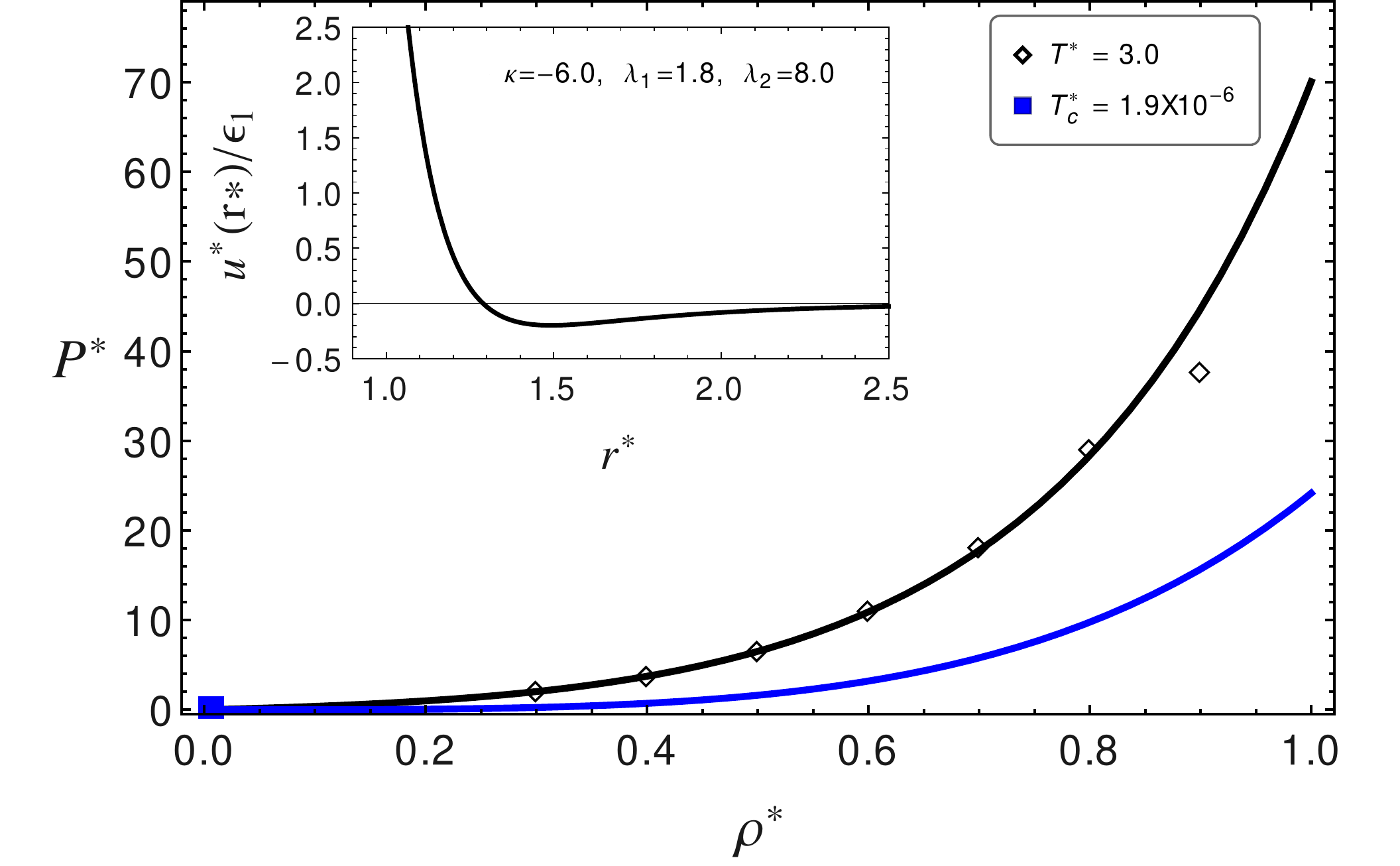}\\
	\includegraphics[height=5cm]{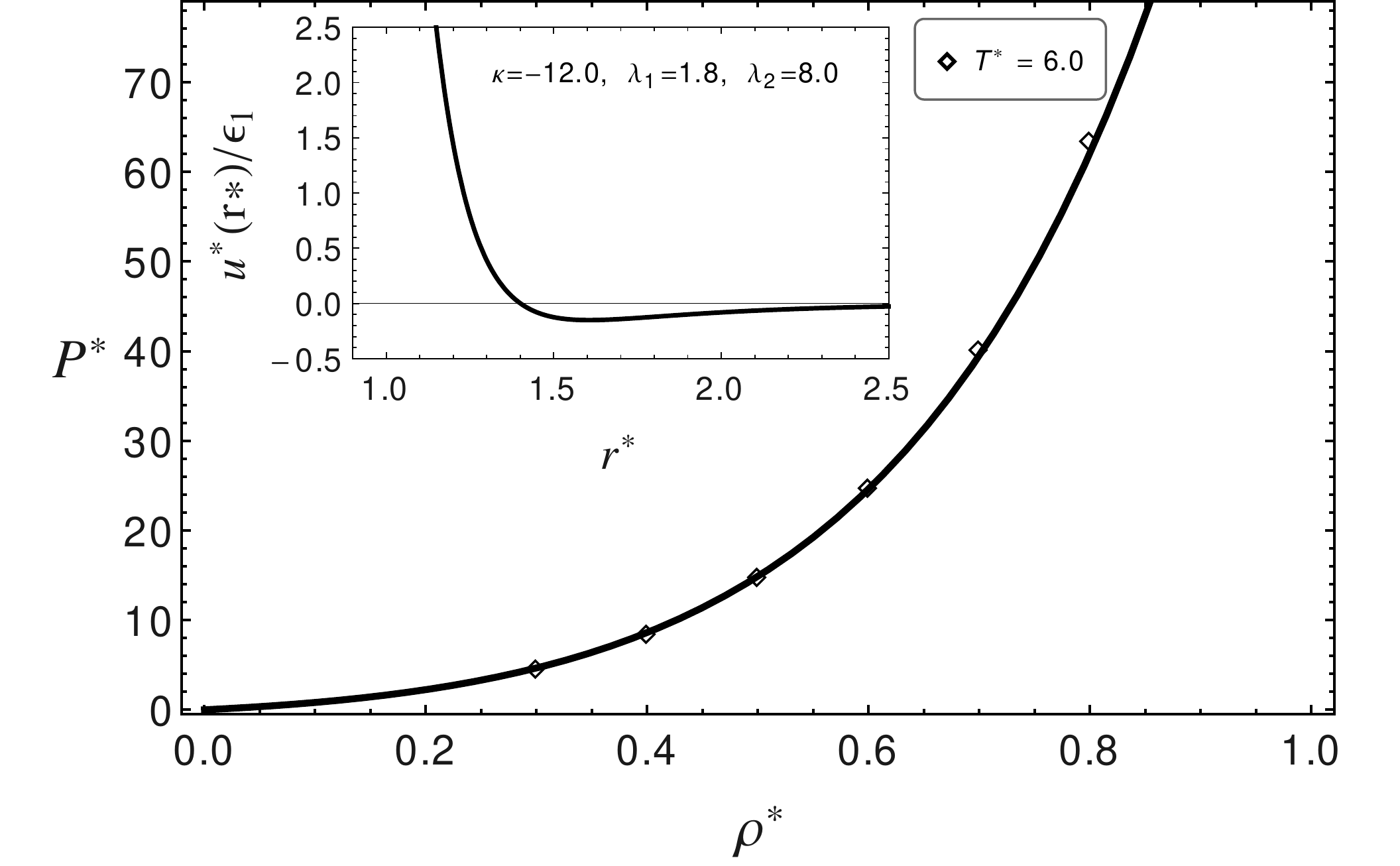}
	\includegraphics[height=5cm]{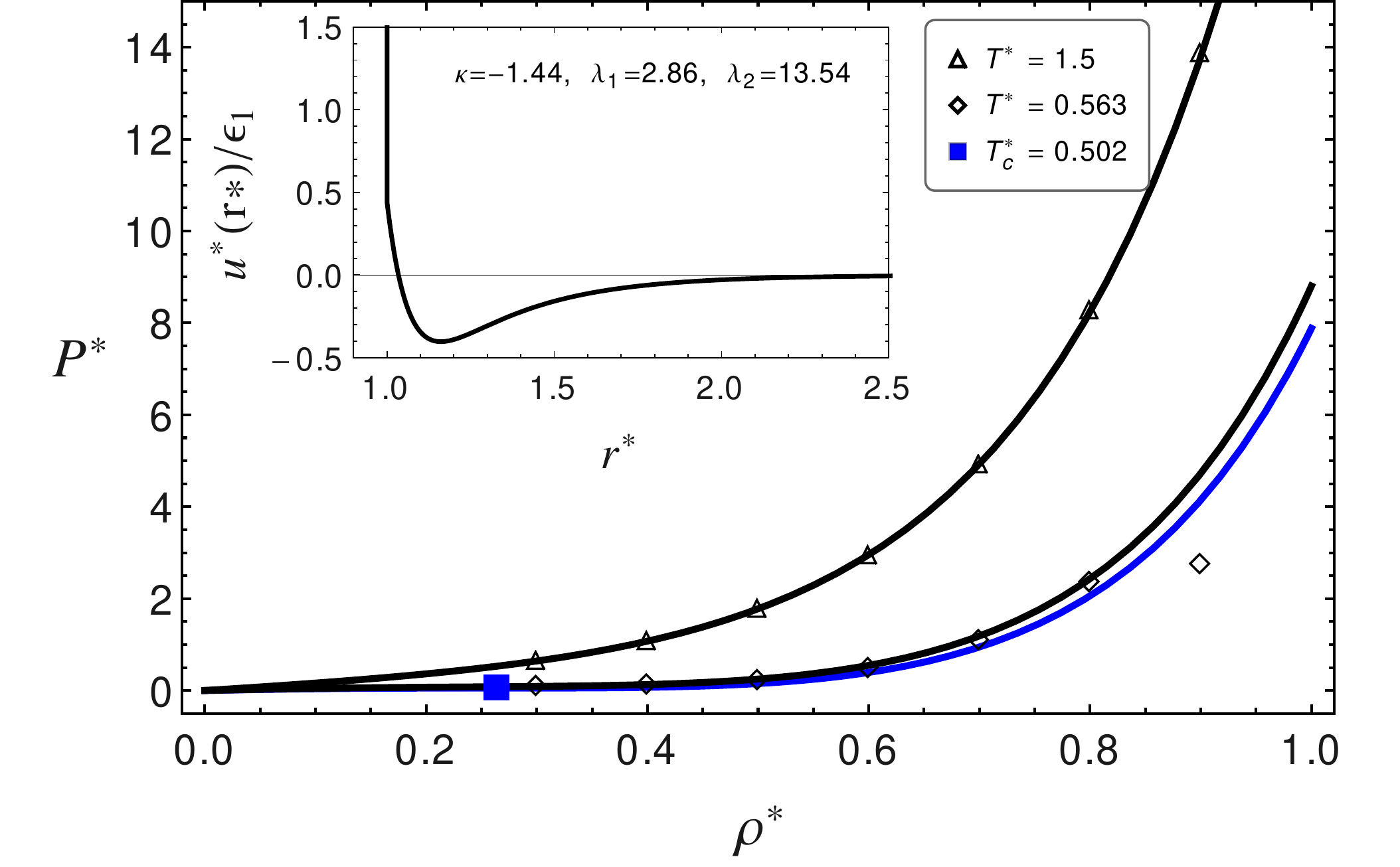}
	\caption{Equations of state for the HCDY fluid computed from Eq. (\ref{ZLJ}) and compared with the Montecarlo simulations of Lin {\it et al.} \cite{Lin}. In the inset we include the shape of the potential for each case. The blue lines are the critical isotherms and the blue squares are the critical points predicted with the thermodynamic perturbation theory approach.}	
	\label{Ps}
\end{figure}

The results for the isotherms corresponding to all cases as computed with Eq.\ (\ref{ZLJ}) with $n=2$ are shown in Fig.\ \ref{Ps}, along with the simulation data. As clearly seen, the agreement between theory and simulation is very satisfactory and equivalent to the performance of the results derived from the EOS of Lin {\it et al.} \cite{Lin} (not shown). For cases 1, 2 and 6 we also include our results for the critical isotherms together with the corresponding critical points. These were not reported in Ref.\ \cite{Lin} but the overall features that we get for such isotherms seem to be consistent with the rest of the simulation data.

\begin{figure}
	\begin{center}
	\includegraphics[height=7cm]{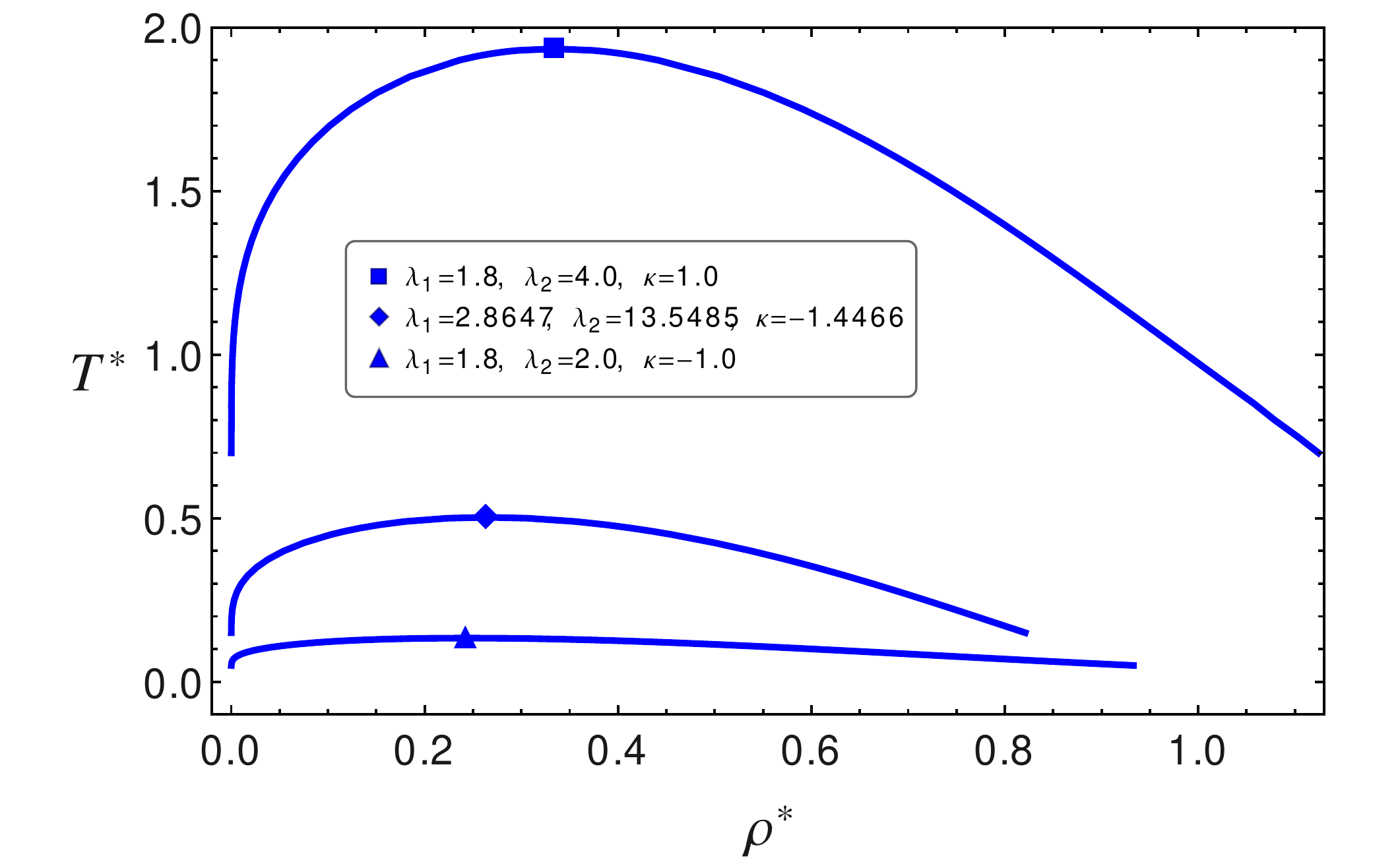}
	\end{center}
	\caption{Liquid-vapor coexistence curves for three different HCDY fluids computed from Eq.\ (\ref{ZLJ}). The blue squares are the predicted critical points.}	
	\label{Ps2}
\end{figure}

Next we concentrate on the liquid-vapor coexistence curves for those cases (1, 2 and 6) where the HCDY fluid has a critical point. These have been derived directly from Eq.\ (\ref{ZLJ}) by requiring the usual conditions of equal pressures and chemical potentials of both phases. The results are shown in Fig.\ \ref{Ps2} where the numerically estimated critical points have also been included. The values of the (reduced) critical constants $\rho^*_c= 6/\pi \eta_c$, $T^*_c$ and $P^*_c= P_c d^3/\epsilon_1$ for these cases are contained in Table \ref{table1}.
Note that numerically we get a physical critical point for case 4, whose critical values are also included in Table \ref{table1}. Since these critical values are very  close to zero, we can not be certain that such a critical point is physical or an artifact of the numerical evaluation, since in this case we were not able to compute the corresponding coexistence curve. Finally, and as far as we are aware, there are no simulation data for the binodals of cases 1, 2 and 6 and so a comparison was not possible in this instance.

\begin{table}
	\begin{center}
			\caption{Values of the critical constants for cases 1, 2, 4 and 6.}
			\label{table1}
		\begin{ruledtabular}
		\begin{tabular}{cccc}
			& $\rho^*_c$ & $T^*_c$ & $P^*_c$ \\
			Case 1 & 0.334612 & 1.93487 & 0.268408 \\
			Case 2& 0.243225 & 0.133628 & 0.011984 \\
			Case 4 & 0.0066769 & 0.000852366 & 1.90427$\times 10^{-6}$ \\
			Case 6 & 0.264336 & 0.502825 & 0.0537199 \\
			
		\end{tabular}	
			\end{ruledtabular}
	\end{center}

\end{table}

\subsection{Stochastic analysis of critical points}

\begin{figure}
	\begin{center}
		\includegraphics[height=6cm]{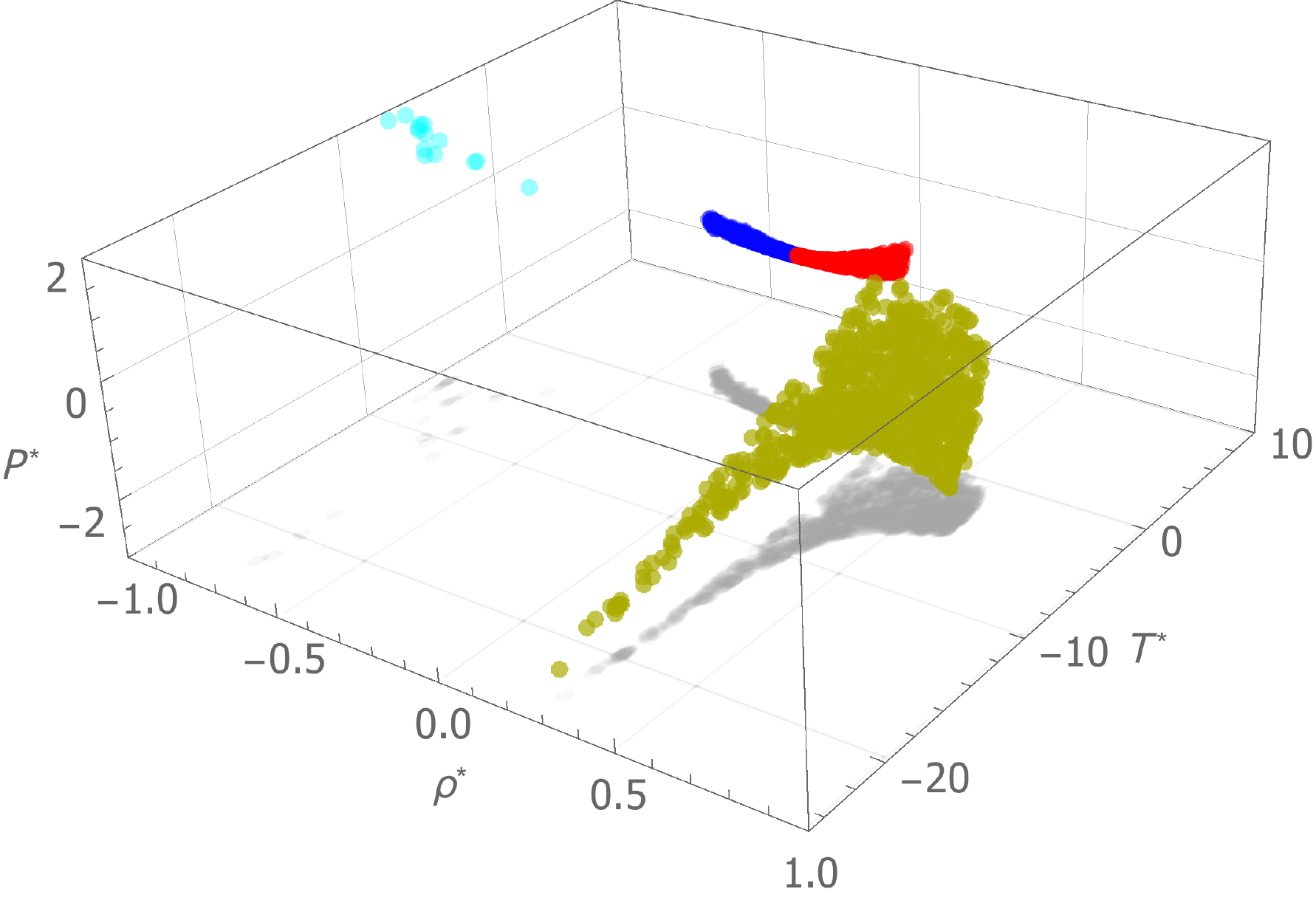} \hspace{1cm}
		\includegraphics[height=6.5cm]{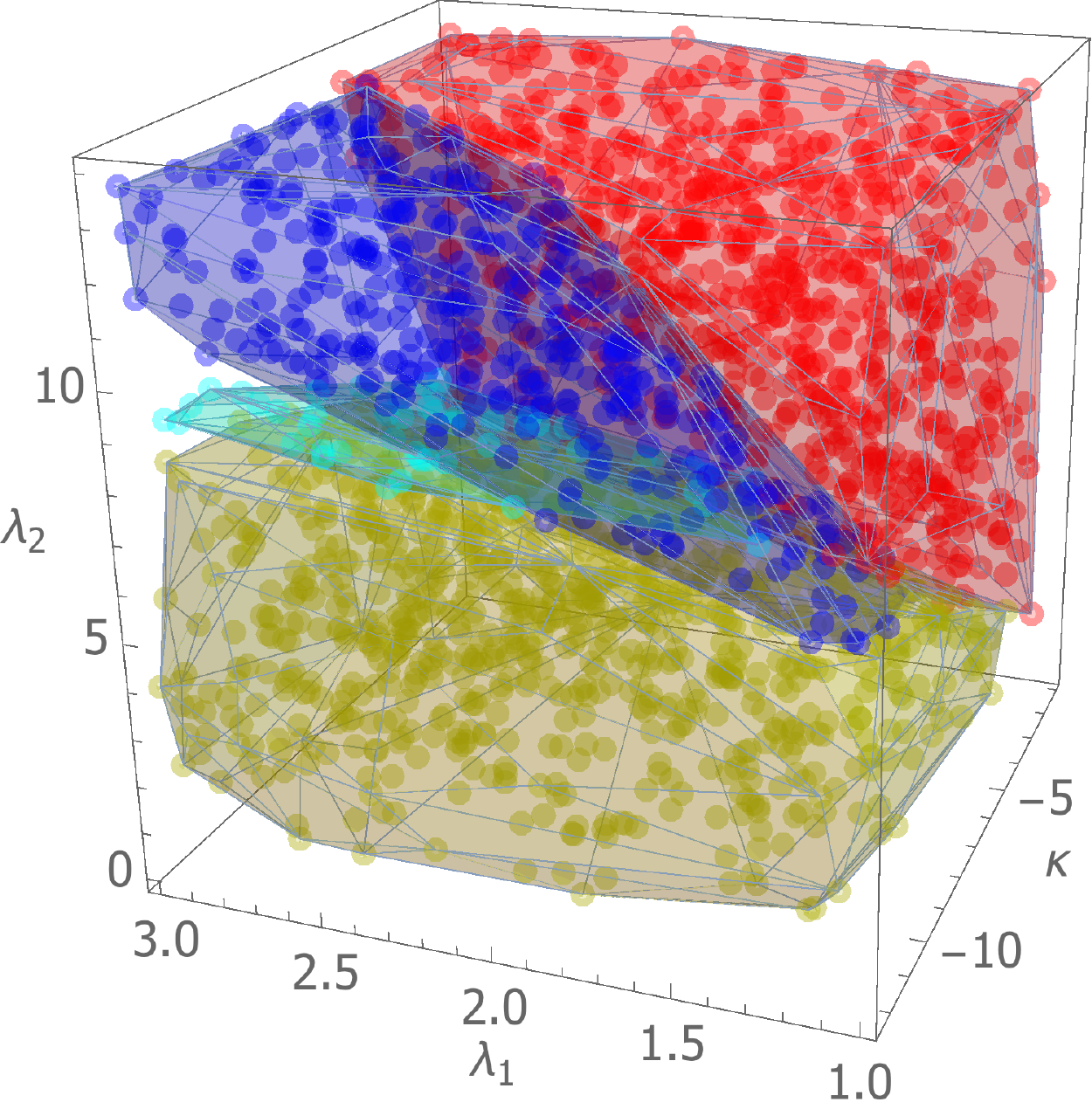}\\
		(a)\hspace{7.5cm}(b)\\
		\caption{(a) Critical points obtained from the theoretical scheme. Red points are in the physical region. Yellow, blue and cyan points are in different octants in the $\rho^*$, $T^*$ and $P^*$ space, where at least one of the critical constants has a negative value. (b) Examined space of parameters $\kappa$, $\lambda_1$ and $\lambda_2$. The colors are mapped from the different regions found in the critical points computation. A convex hull around each set of points has been also included to aid in the visualization of the boundaries of such regions. }
		\label{Pcrit}
	\end{center}
	
\end{figure}

The results of the previous subsection confirm the usefulness of counting with a relatively simple (completely analytical) EOS for the HCDY fluid. However, this fact may be further exploited to investigate in particular the effect of the values of the different parameters ($\lambda_1$,$\lambda_2$ and $\kappa$)  on the liquid-vapor critical point. To this end we sampled 2000 points in the  space spanned by the values of the parameters in the following intervals $\kappa \in [-13,-1]$, $\lambda_1 \in [1,3]$ and $\lambda_2 \in [0,18]$. These intervals embody five of the previous six cases (except for case 1) and allow us to probe the most significant region. As noted above, for some values of the parameters no critical point will occur. Nevertheless, we let Eq. (\ref{ZLJ}) produce 'critical' points with and without physical meaning and sort them in the octants of the 3D space of (critical) density $\rho^*_c$, temperature $T^*_c$ and pressure $P^*_c$. Only those critical points where the three values turn out to be  positive, are physically acceptable. The other cases may be related with different (particular) shapes of the potential. The results for the critical points are displayed in Fig.\ \ref{Pcrit}(a).  The red points correspond to the physically acceptable critical points. The different regions can in turn be mapped onto the parameter space spanned by the values of $\kappa$, $\lambda_1$ and $\lambda_2$. This is shown in fig. \ref{Pcrit}(b). Again the red points correspond to the values of the parameters that will lead to the appearance of a physical critical point.

\begin{figure}
	\begin{center}
		\includegraphics[height=7.5cm]{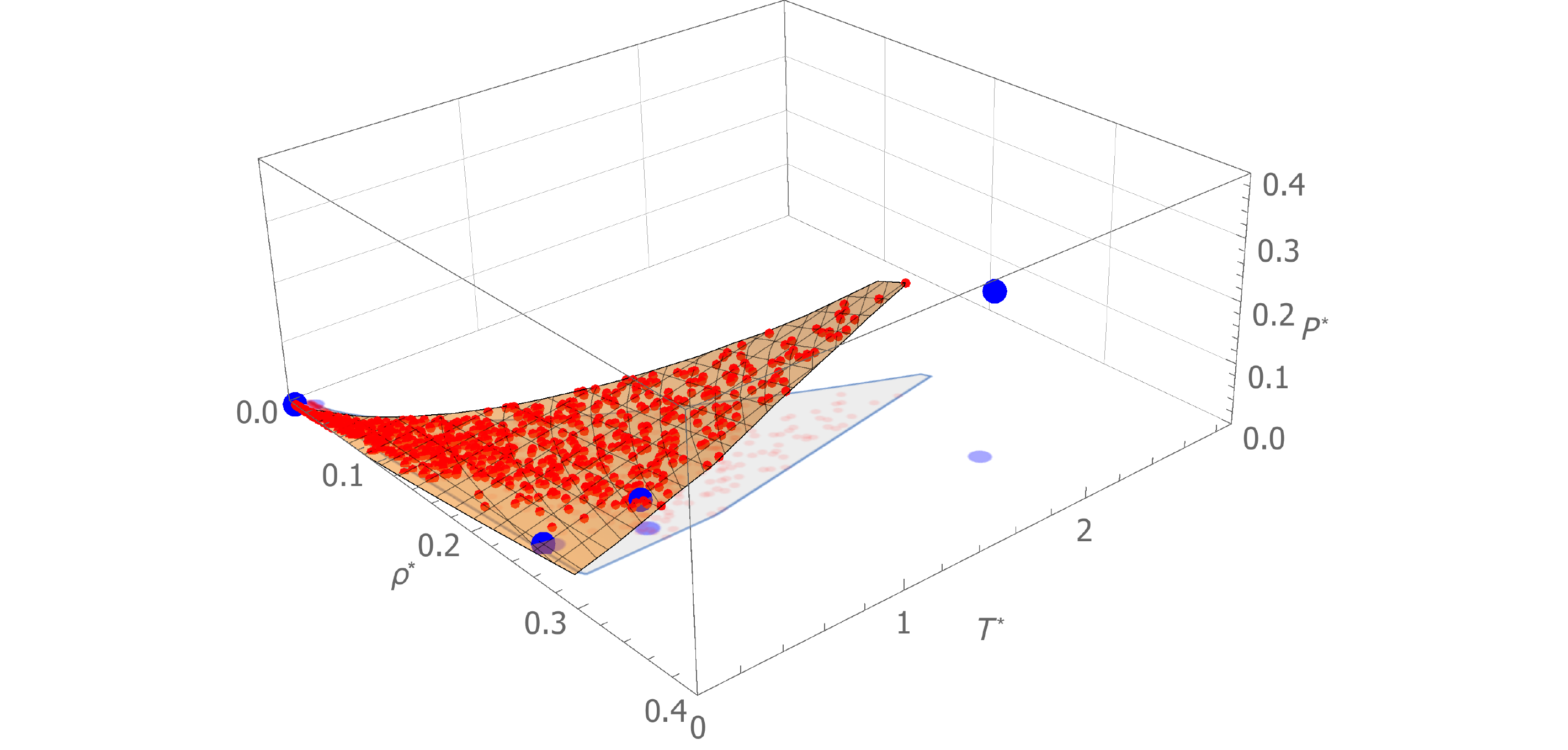}
		\caption{Critical points with physical sense obtained from Eq.\ (\ref{ZLJ}). The surface included is a rough first approximation to the region where physical solutions may be obtained from the sample space proposed. Blue points correspond to the four cases examined by Lin {\it et al.}\cite{Lin} in which we found critical points.}
		\label{preal}
	\end{center}
\end{figure}

A deeper view of the region with physical meaning is in Fig. \ref{preal}. There we have included an approximate surface that bounds the cloud of points to add a reference to the three dimensional view. Also the critical points for cases 1, 2, 4 and 6 have been included, even  when one of them (case 1) falls outside the range of chosen values for the parameters since $\kappa$ is positive in that case. We should remark that critical points tend to cluster in a sharp corner close to the origin of the thermodynamic variables, and they fall more frequently there than in the other borders.

\begin{figure}
	\begin{center}
		\includegraphics[height=7.5cm]{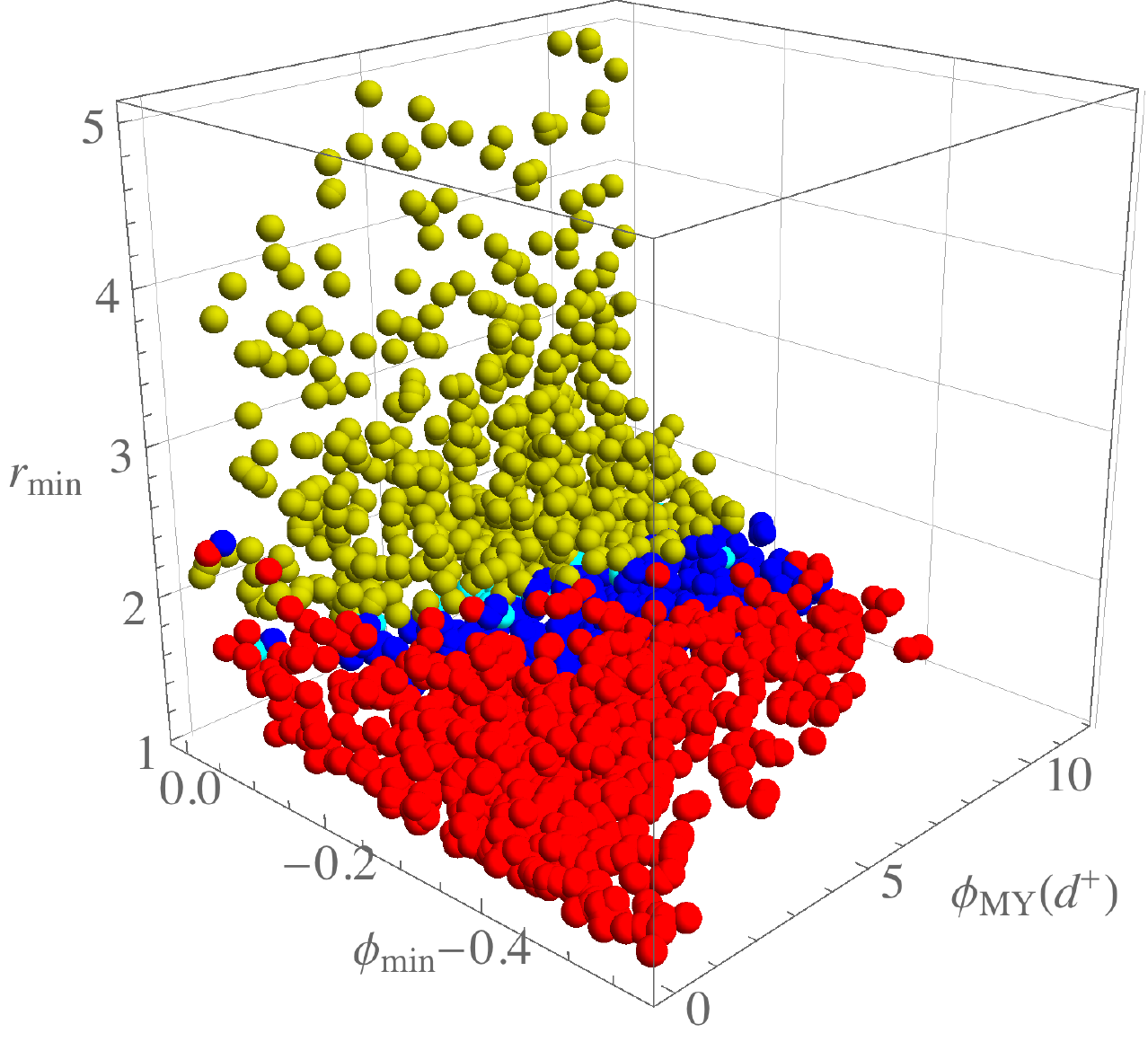}
		\caption{Shape parameters $\phi_{min}$, $\phi_{MY}(d^+)$ and $r_{min}$  for all sampled points. Again the color maps the regions of physical meaning described in the critical points calculation.}
	\end{center}
	\label{Potential}
	
\end{figure}

While the above mapping is already illustrative of the connection between the values of the parameters in the potential and the critical behavior of the fluid, one may also make the connection between the geometry of the potential and this critical behavior. To this end we consider the contact value of the potential $\phi_{DY}(d^+)$, the position of the minimum $r_{min}$ and the value of the potential well at such minimum $\phi_{min}=\phi_{DY}(r_{min})$. In particular, if $\phi_{DY}(d^+) >0$ then the potential contains a repulsion outside the core, as in cases 3 -- 6. The graphical representation of this connection is shown in Fig.\ \ref{Potential}. In the figure we have not included cases where the potential does not have a well. Such cases appear in a few sampled points and lead to non physical solutions for the critical points. A relevant observation of this graph is that the obtained equation of state gives non physical solutions for some potentials with a small but real well. When the well depth becomes  small and the position of the minimum rises, the critical points approach very fast the origin of the thermodynamic variables and then some of them become unphysical. This effect seems to have a very small dependence on the value of  $\phi_{DY}(d^+)$. Although interesting, such result must be taken with caution since it may well be a consequence of the restriction to a first order thermodynamic perturbation theory and not a real physical effect.

\section{Summary and concluding remarks}
\label{Conclu}

In this paper we used the first order in the inverse temperature thermodynamic perturbation theory to derive a completely analytical equation of state for the HCMY fluid. As compared with previous work on this subject, the new aspect concerns the consideration of the structural properties of the hard-sphere fluid as given by the RFA approach.\cite{Bravo1,Bravo2,HYS08} This latter has as a particular asset that the resulting structure is thermodynamically consistent and requires as its only input an equation of state for the hard-sphere fluid.

The fact that we have a closed-form analytical equation of state allows for the examination of the effect of the values of the potential parameters on the thermodynamic properties of the system. Restricting ourselves to the HDCY fluid we have been able to compute isotherms, liquid-vapor coexistence curves and the critical constants (these last two properties when they exist) for the six systems that were previously considered by Lin {\it et al.}\cite{Lin} Both the computation of the critical points and the coexistence curves are to our knowledge completely new and call for confirmation from simulation. We hope that our results will stimulate the performance of such simulations.
Nevertheless, based on the information pertaining to the six cases, we performed an stochastic analysis of the potential parameters leading to a physical liquid-vapor critical point. It turns out that, within the present approximation, one may clearly identify regions in the parameters space where the corresponding potential will not produce a liquid-vapor phase transition. Although this evidence can not be considered as wholly conclusive given the approximations involved, it may certainly serve as a guide for further simulations or for the application of the HCDY fluid to model a specific system.

Finally, we should mention that a similar development can be carried out for a multi-Yukawa potential without a hard core. This only requires the use of an 'effective' hard-core diameter for which specific recipes exist within the thermodynamic perturbation theory\cite{BH67,Mansoori1,Mansoori2,RS,WCS71} in which the hard-sphere fluid is taken as the reference system. While we have performed some calculations following this strategy, the simplicity of a totally analytical development is lost and the actual computations become much more involved. We nevertheless plan to address this point in more depth in future work.


\begin{thebibliography}{99}
	\bibitem{McQuarrie0}See for instance D. A. McQuarrie, \textit{Statistical Mechanics} (Harper \& Row, New York, 1976) or any other textbook in Statistical Mechanics.
	
	\bibitem{JM80} C. Jedrzejek and G. A. Mansoori, Acta Physica Polonica \textbf{A57}, 107 (1980).
	
	\bibitem{GS83} D. J. Gonz\'alez and M. Silbert, J. Phys. C: Solid State Phys. \textbf{16}, L1097 (1983).
	
	\bibitem{KJ88} J. Konior and C. Jedrzejek, Mol. Phys. \textbf{63}, 655 (1988).
	
	\bibitem{RC89} E. N. Rudisill and P. T. Cummings, Mol. Phys. \textbf{68}, 629 (1989).
	
	\bibitem{R89} J. S. Rowlinson, Physica A \textbf{156}, 15 (1989).
	
	\bibitem{R93} Y. Rosenfeld,  J. Chem. Phys. \textbf{98}, 8126 (1993).
	
	\bibitem{KC96} Yu. V. Kalyuzhnyi and P. T. Cummings, Mol. Phys. \textbf{87}, 1459 (1996).
	
	\bibitem{Tang:1997} Y. Tang, Z. Tong and B. C.-Y. Lu, Fluid Phase Equilib. \textbf{134}, 21 (1997).
	
	\bibitem{AOAS99} I. Ali, S. M. Osman, M. Al-Busaidi and R. N. Singh, Int. J. Mod. Phys. B \textbf{13}, 3261 (1999).
	
	\bibitem{AOS99} I. Ali , S.M. Osman, R.N. Singh, J. of Non-Cryst. Solids \textbf{250-252}, 364 (1999).
	
	\bibitem{BU00} L. Blum and M. Ubriaco, Mol. Phys. \textbf{98}, 829 (2000).
	
	\bibitem{Lin}Y.-Z. Lin, Y.-G. Li, J. F. Lu and W. Wu, J. Chem. Phys. \textbf{117}, 22 (2002).
	
	\bibitem{S03} J. X. Sun, Phys. Rev. E \textbf{68}, 061503 (2003).
	
	\bibitem{S04} J. X. Sun, Chemical Physics \textbf{302}, 203 (2004).
	
	\bibitem{Lin04}Y.-Z. Lin, Y.-G. Li and J. F. Lu, Mol. Phys. \textbf{102}, 63 (2004).
	
	\bibitem{G04} H. Gu\'erin, Fluid Phase Equilib. \textbf{218}, 47 (2004).
	
	\bibitem{PA05} Y. Pathania and P. K. Ahluwalia, Pramana \textbf{65}, 457 (2005).
	
	\bibitem{Lin06} Y.-Z. Lin, Y.-G. Li and J.-D. Li, J. Mol. Liquids \textbf{125}, 29 (2006).
	
	\bibitem{PA06} Y. Pathania and P. K. Ahluwalia, Pramana \textbf{67}, 1141 (2006).
	
	\bibitem{K06} M. Bahaa Khedr, Int. J. Mod. Phys. B \textbf{20}, 3373 (2006).
	
	\bibitem{AW07} A. J. Archer and N. B. Wilding, Phys. Rev. E \textbf{76}, 031501 (2007).
	
	\bibitem{CBLFB07} S.-H. Chen, M. Broccio, Y. Liu, E. Fratinid and P. Baglioni, J. Appl. Cryst. \textbf{40}, s321 (2007).
	
	\bibitem{APER07} A. J. Archer, D. Pini, R. Evans, and L. Reatto, J. Chem. Phys. \textbf{126}, 014104 (2007).
	
	\bibitem{YJ08} Y.-X. Yu and L. Jin, J. Chem. Phys. \textbf{128}, 014901 (2008).
	
	\bibitem{AIPR08} A. J. Archer, C. Ionescu, D. Pini and L. Reatto, J. Phys.: Condens. Matter \textbf{20}, 415106 (2008).
	
	\bibitem{LHSRWB10} L. L. Lee, M. C. Hara, S. J. Simon, F. S. Ramos, A. J. Winkle, and J.-M. Bomont, J. Chem. Phys. \textbf{132}, 074505 (2010).
	
	\bibitem{BBC10} J.-M. Bomont, J.-L. Bretonnet, and D. Costa, J. Chem. Phys. \textbf{132}, 184508 (2010).
	
	\bibitem{KNMT10} J. Krej\v{c}\'{\i}, I. Nezbeda, R. Melnyk, and A. Trokhymchuk J. Chem. Phys.  \textbf{133}, 094503 (2010).
	
	\bibitem{KPH10} M. Khanpour, G.A. Parsafar and  R. Hashim, J. Non-Cryst. Solids \textbf{356}, 2247 (2010).
	
	\bibitem{KRMT11} J. Krej\v{c}\'{\i}, I. Nezbeda, R. Melnyk and A. Trokhymchuk, Condens. Matt. Phys. \textbf{14}, 33005 (2011).
	
	\bibitem{Kim11} J. M. Kim, R. Casta\~{n}eda-Priego, Y. Liu, and N. J. Wagner, J. Chem. Phys. \textbf{134}, 064904 (2011).
	
	\bibitem{FH11} N. Farzi and S. Hoseini, Chem. Phys. \textbf{384}, 9 (2011).
	
	\bibitem{OAS13} S. M. Osman, I. Ali, and R. N. Singh, Phys. Rev. E \textbf{87}, 012122 (2013).
	
	\bibitem{H15} S. Hslushak, J. Chem. Phys. \textbf{143}, 124906 (2015).
	
	\bibitem{BH67} J. A. Barker and D. Henderson, J. Chem. Phys. {\bf 47}, 2856 (1967); {\it ibid} {\bf 47}, 4714 (1967).
	
	\bibitem{Mansoori1}  G. A. Mansoori and F. B. Canfield, J. Chem. Phys. {\bf 51}, 4958 (1969).
	
	\bibitem{Mansoori2}  G. A. Mansoori, J. A. Provine and F. B. Canfield, J. Chem. Phys. {\bf 51}, 5295 (1969).
	
	\bibitem{RS}  J. Rasaiah and G. Stell, Mol. Phys. {\bf 18}, 249 (1970).
	
	\bibitem{WCS71} J. D. Weeks, D. Chandler and H. C. Andersen, J. Chem. Phys. {\bf 54}, 5237 (1971).
	
	\bibitem{Bravo1}  S. Bravo Yuste and A. Santos , Phys. Rev. A {\bf 43}, 5418 (1991).
	
	\bibitem{Bravo2}  S. Bravo Yuste, M. L\'{o}pez de Haro and A. Santos, Phys. Rev. E {\bf 53}, 4820 (1996).
	
	\bibitem{HYS08}
	M. L\'opez de Haro, S. B. Yuste and A. Santos, \emph{Alternative approaches to the equilibrium properties of hard-sphere liquids}, in {\em Theory and Simulation of Hard-Sphere Fluids and Related Systems, in the series Lecture Notes in Physics, Vol.\ \textbf{753}}, edited by A. Mulero (Springer, Berlin, 2008), pp. 183--245.
	
	\bibitem{wertheim}  M. S. Wertheim,  Phys. Rev. Lett. {\bf 10}, 321 (1963).
	
	\bibitem{CS}  N. F. Carnahan and K. E. Starling, J. Chem. Phys. {\bf 51}, 635 (1969).
	
	
	
\end{thebibliography}
\end{document}